# Effect of nonlinearity in the pass-through optics on femtosecond laser filament in air

Alexander A. Dergachev,[1] Andrey A. Ionin,[2] Valery P. Kandidov,[1] Daria V. Mokrousova,[2,3]
Leonid V. Seleznev,[2] Dmitry V. Sinitsyn,[2] Elena S. Sunchugasheva,[2,3,*]
Svyatoslav A. Shlenov,[1] Anna P. Shustikova[2,3]

[1] *Department of Physics and International Laser Center, M.V. Lomonosov Moscow State University, Vorob'evy gory, 119991 Moscow, Russia*
[2] *P.N. Lebedev Physical Institute of Russian Academy of Sciences, 53 Leninskiy prospect,119991 Moscow, Russia*
[3] *Moscow Institute of Physics and Technology, 9 Institutskiy Pereulok, 141701 Dolgoprudny, Moscow Region, Russia*
*Corresponding author: ses@lebedev.ru*

*An influence of pass-through optics on femtosecond laser pulse filamentation in ambient air is analyzed for the first time both experimentally and numerically. Propagation of high-power femtosecond laser pulse through solid optical elements introduces spatiotemporal phase modulation due to the Kerr effect. This modulation may have a strong effect on the pulse filamentation in air. We demonstrated that the phase modulation obtained in the thin pass-through dielectric plate reduces the distance to the filament onset and increases the plasma channel length*

Femtosecond laser pulse filamentation is the result of nonlinear beam self-focusing in the medium, defocusing in self-induced laser plasma, diffraction and dispersion [1-3]. Filamentation takes place when the peak power of the pulse exceeds the critical power of self-focusing $P_{cr}$. For Gaussian beam

$$P_{cr} = \frac{3.77\lambda^2}{8\pi n_0 n_2},$$

where $\lambda$ is the central wavelength and $n_2$ is nonlinear coefficient of the medium. In gases $P_{cr}$ is about tenths to several gigawatts, in condensed matter the coefficient $n_2$ is much higher and critical power decreases to several megawatts. For the 800 nm wavelength in atmospheric air $n_2 \approx (2.4 \div 5.8) \cdot 10^{-19}$ cm$^2$/W [4-5] and critical power $P_{cr} \approx 1.7 \div 4$ GW [6], in fused silica $n_2 \approx (2.4 \div 3.2) \cdot 10^{-16}$ cm$^2$/W and $P_{cr} \approx 2 \div 4$ MW [7], in CaF$_2$ $n_2 \approx 1.24 \cdot 10^{-16}$ cm$^2$/W and $P_{cr} \approx 7$ MW [8].

High values of nonlinear coefficient $n_2$ in solid dielectrics impose strict limitations on the usage of pass-through optical elements in the high-power femtosecond laser systems. For TW systems reflecting optics is used to prevent nonlinear distortions of the laser beam and damage in pass-through elements [9]. At the same time, some necessary elements of the high-power femtosecond systems, like windows of gas cuvettes, vacuumed compressors, frequency mixers, attenuators, etc. may seriously corrupt the beam. In [10] small-scale intensity fluctuations in cross-section of the beam increase significantly after passing through the output CaF$_2$ window of the compressor 1 cm in thickness.

In high-power ultrafast laser amplifiers the influence of the Kerr nonlinearity is usually estimated by B-integral

$$B = \frac{2\pi}{\lambda} \int_0^{\Delta z} n_2 I(z) dz,$$

which determines the total phase shift on the beam axis after passing the distance $\Delta z$ [11]. If $B > 1 \div 2$ the beam may break down due to small-scale self-focusing in transparent components of the amplifier. Initial intensity of about $10^{11}$ W/cm$^2$, common for air filamentation experiments [1], leads to $B < 1$ for transparent dielectrics less than 1 cm in thickness. The breakdown therefore does not take place. Nevertheless, the nonlinear self-phase modulation obtained by the pulse in this case may significantly affect its filamentation on the long distance in air. Likewise direct phase modulation [12] or amplitude modulation [13], nonlinear phase modulation affects plasma channels generated during high-power pulse propagation.

In this paper, we studied the effect of the Kerr nonlinearity in thin fused silica plane plates on the femtosecond laser pulse filamentation in air. We demonstrated the significant change in the plasma channels length caused by the plates of various thicknesses to take place.

First, consider propagation of the laser pulse through a transparent optical element. Because of the Kerr effect laser pulse, propagating through transparent optical element, obtains self-phase modulation $\Delta\varphi(r, t)$ depending from both radial and time coordinates. Assuming that the element's thickness $\Delta z$ is sufficiently small we may neglect the change in the pulse intensity $I(r, t)$. In this case the phase modulation is given by the following formula:

$$\Delta\varphi(r,\tau) = -n_2^{solid} I(r,\tau) \frac{2\pi n_0}{\lambda} \Delta z, \quad (1)$$

where $I(r, t)$ is intensity distribution in the pulse, $n_2^{solid}$, $n_0$, $\Delta z$ are nonlinearity coefficient, refractive index and thickness of the pass-through element.

Phase modulation $\Delta\varphi(r, t)$ in the pulse leads to time-dependent or dynamic wavefront change of the laser beam, i.e. different wavefront change in different pulse time slices. For Gaussian beam

$$I(r,\tau) = I_0(\tau)\exp\{-r^2/r_0^2\}$$

we can estimate time-dependent on-axis curvature of the wavefront or the equivalent focal length of the Kerr lens $f_{nl}(\tau)$ formed by the pass-through element:

$$f_{nl}(\tau) = \frac{r_0^2}{2n_2^{solid} I_0(\tau) n_0 \Delta z} \text{ or } f_{nl}(\tau) = \frac{4\pi^2 r_0^4}{3.77\lambda^2 \Delta z} \frac{P_{cr}^{solid}}{P(\tau)}, \quad (2)$$

where $I_0(\tau)$ is on-axis pulse intensity, $P_{cr}^{solid}$ is critical power of self-focusing in the element, $P(\tau)$ is pulse power in different time slices and $r_0$ is beam radius at $e^{-1}$ level. If the pulse is initially focused by a

mirror with focal length $f_0$, the total focal length $f(\tau)$ can be estimated as

$$f^{-1}(\tau) = f_0^{-1} + f_{nl}^{-1}(\tau). \quad (3)$$

According to the formula (1) 100 fs Gaussian pulse at wavelength 744 nm with energy 2.2 mJ and Gaussian beam profile with radius $r_0 = 1.6$ mm propagating through 10 mm fused silica plate acquires dynamic on-axis wavefront curvature with minimal focal length 130 cm. This value is comparable to geometrical focusing in laboratory experiments [14]. The focus distance attains minimum in the central time slices of the pulse when the power reaches its peak value 20 GW (about 10 $P_{cr}$ in air). At the leading and trailing edges of the pulse where power is less than in the central part nonlinear focusing is weaker. The time dependence of focal length $f(\tau)$ during the pulse for two fused silica plates is shown in Fig. 1. For the parameters given above the minimal focal lengths are 122 cm and 87 cm for the plates of 4.5 mm and 10.5 mm in thickness, correspondingly.

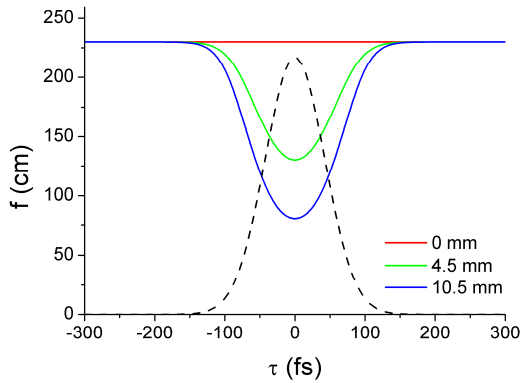

Fig. 1 Focal distance $f(\tau)$ in different time slices of the Gaussian pulse with peak intensity $2.3 \cdot 10^{11}$ W/cm$^2$ and geometrical focusing $f_0 = 230$ cm (marked by red line). Dynamic wavefront curvature is introduced by a thin fused silica plate of 4.5 and 10.5 mm in thickness. Dashed line schematically represents the intensity profile of the laser pulse.

Numerical simulation of the femtosecond laser pulse filamentation with dynamic wavefront was based on the system of equations for slowly varying envelope of electric field and free electrons concentration. The system takes into account diffraction, group velocity dispersion, Kerr and plasma nonlinearities and photoionization of air [14].

We studied filamentation of the Gaussian pulse with Gaussian beam profile after passing through fused silica plates of different thickness. Initial pulse envelope immediately after the plate was set as follows:

$$A(r, \tau, z=0) = A_0 \exp(-\frac{\tau^2}{2\tau_0^2}) \exp(-\frac{r^2}{2r_0^2}) \exp(i\frac{kr^2}{2f_0} + i\Delta\varphi(r,\tau)) \quad (4)$$

where $\Delta\varphi(r,t)$ is self-phase modulation in the plate according to (1). After numerical solution of the self-consistent system for field envelope and free electrons concentration we calculated fluence

$$F(r, z) = \int I(r, \tau, z) d\tau$$

and plasma concentration $n_e(r,z)$ immediately after the pulse. Also we calculated linear plasma concentration $\rho_{lin}(z)$ which was measured in the experiment [14]:

$$\rho_{lin}(z) = \int n_e(r,z) \cdot 2\pi r dr. \quad (5)$$

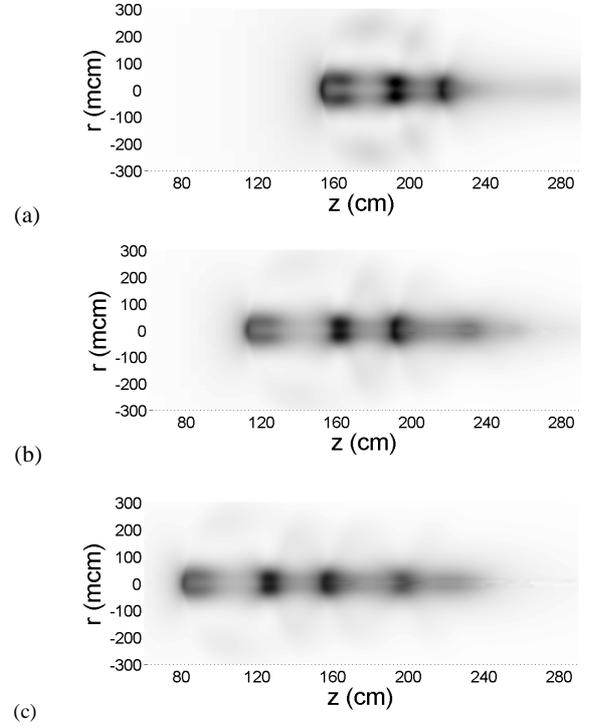

Fig. 2 Fluence distributions F(r, z) along the propagation axis for various plates: (a) $\Delta z = 0$ mm, (b) $\Delta z = 4.5$ mm, (c) $\Delta z = 10.5$ mm. Geometrical focusing $f_0 = 230$ cm. Dark regions correspond to higher fluence.

Figure 2 shows spatial distributions of fluence F(r, z) in the pulse obtained in numerical simulation. It is clearly seen that nonlinear phase modulation of the pulse obtained in the fused silica plate leads to the increase in the filament length. Due to the greater on-axis wavefront curvature the central time slices of the pulse are focused closer to the focusing mirror thus stimulating earlier filament and plasma channel onset (Fig. 3, solid lines). The slices on the leading edge of the pulse which have smaller wavefront curvature are focused at about the same distance as without any plates.

Dark regions in fluence distributions (Fig. 2) and the peaks in the distribution of linear plasma concentration (Fig. 3, solid lines) correspond to the pulse refocusing during filamentation. The number of refocusings increases with plate thickness. Additional peaks of the concentration at the end of the plasma channel in the presence of the plate are the cooperative result of refocusing of the trailing edge of the pulse and self-focusing of its leading edge. Thus in the presence of the plate the plasma channel starts earlier than without a plate but ends at practically the same distance due to these additional peaks. The channel therefore undergoes elongation. In addition, the total number of free electrons in generated plasma increases from slightly $8 \cdot 10^{13}$ cm$^{-1}$ without a plate to $9.5 \cdot 10^{13}$ cm$^{-1}$ with 10.5 mm plate.

A series of experiments were performed to verify influence of pass-through optics on plasma column of femtosecond laser filament in air. In our experiments we used Ti:Sapphire laser system [15] which produced 100 fs FWHM pulses with energy 2.2 mJ at the wavelength of 744 nm. The laser pulses were focused with 3-m spherical mirror. In front of the mirror at the distance of 70 cm off it fused silica plates 4.5 mm or 10.5 mm in thickness were placed. Another thin probe glass plate moving along the propagation axis was used to determine the filamentation onset [16]. When the probe plate was near the filament onset, filamentation started in the plate with clearly observable supercontinuum generation. Taking into account the ratio of air and glass nonlinearity coefficients, we had to add the probe plate thickness increased by three orders of magnitude to the registered position of the plate. We used 0.15 mm probe plate, so the accuracy of the filamentation onset measurement was about 15 cm.

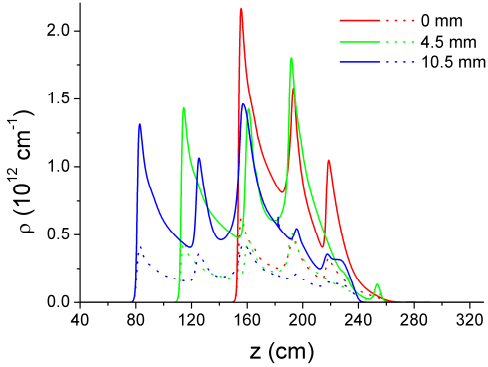

Fig. 3 Linear plasma concentration calculated in numerical simulation along the optical axis immediately after the laser pulse $\rho_{lin}(z)$ (solid lines) and concentration averaged over the registration time $\rho_{ave}(z)$ (dotted lines) for plates of various thickness: $\Delta z = 0$ mm (red line), 4.5 mm (green line), 10.5 mm (blue line).

For plasma concentration measurements we used two metal hemispherical electrodes 2 cm in diameter placed at the distance of 3 mm. The voltage between them was 300 V. Laser pulse propagated through the space between the electrodes [14]. When plasma was generated by the propagating pulse, conductivity between the electrodes changed and we detected recharging current and variation of voltage by an oscilloscope. Longitudinal translation of the electrodes system along the optical axis allowed us to produce measurements of relative plasma concentration integrated over laser beam cross-section (i.e. linear plasma concentration) at different distances from the focusing mirror. The typical response time of the electrodes system and oscilloscope $t_{reg}$ was about 2 ns. The detected signal was proportional to plasma concentration averaged over $t_{reg}$ taking into account free electron relaxation. The evolution of free electrons concentration (plasma concentration) $n_e(r,t,z)$ and positive ions concentration $n_e(r,t,z)$ during relaxation in the plasma channel was calculated on the basis of the following evolution equations:

$$\frac{\partial n_e(r,t,z)}{\partial t} = -\eta n_e(r,t,z) - \beta_{ep} n_e(r,t,z) n_p(r,t,z)$$

$$\frac{\partial n_p(r,t,z)}{\partial t} = -\beta_{ep} n_e(r,t,z) n_p(r,t,z)$$

(6)

where $\eta = 9 \cdot 10^8$ s$^{-1}$ and $\beta_{ep} = 10^{-7}$ cm$^3$/s [17]. System (6) accounts for the attachment of electrons to neutrals with coefficient $\eta$ and electron-ion recombination with coefficient $\beta_{ep}$. Unlike [18] it does not account for comparatively slow processes: impact ionization and ion-ion recombination. Their typical duration is on the order of several microseconds. As the pulse duration is much less than typical relaxation time in laser plasma, we took initial conditions for (6) equal to electron concentration immediately after the pulse:

$$n_e(r,0,z) = n_p(r,0,z) = n_e(r,z).$$

The latter was taken from numerical simulation of the laser pulse filamentation. Finally, we averaged the numerical solution of (6) over $t_{reg}$ and beam cross-section thus obtaining averaged linear concentration of plasma:

$$\rho_{ave}(z) = \frac{1}{t_{reg}} \int_0^{t_{reg}} \int n_e(r,t,z) 2\pi r dr dt \qquad (7)$$

The averaged electron concentration distributions are shown in Fig. 3 by dotted lines. The account for relaxation made distributions smoother but the specific peaks due to pulse refocusing are still clearly seen.

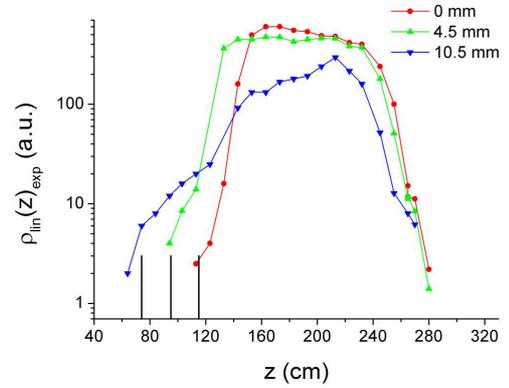

Fig. 4 Experimentally measured relative linear electron concentration $\rho_{lin}(z)_{exp}$ along the optical axis for the plates of various thickness. Vertical lines in the bottom left corner mark the filamentation start indicated by supercontinuum generation in the probe glass plate.

Figure 4 shows the linear concentration in plasma channels measured in the experiment $\rho_{lin}(z)_{exp}$. Filamentation start defined by the thin glass probe plate correlates well with the appearance of the electric signal due to the plasma generation. The results show that nonlinear phase modulation in the fused silica pass-through plates leads to the translation of the channel onset towards the focusing mirror, no displacement of the registered channel end being observed. As a result, the channel length increases for approximately 40 cm (30%) for 10.5 mm plate in comparison with no-plate case. The experimental value of the elongation of plasma channel is in a good agreement with the numerical results.

To summarize, the Kerr nonlinearity in the dielectric pass-through plate leads to spatiotemporal transformation of the femtosecond laser pulse wavefront that significantly influences the pulse filamentation in air. This transformation cannot be simply regarded as an additional nonlinear lens with fixed focal length because different time slices of the laser pulse are self-focused at different distances due to the Kerr-effect driven wavefront curvature. Thin fused silica plate placed into a high-power laser beam decreases a distance to the filamentation onset and increases the plasma channel length and the number of refocusings in the filament. The dynamic wavefront curvature introduced by pass-through optical elements may find applications for the femtosecond filamentation control in gases.

This work was supported by grants: RFBR 14-02-00489, 14-22-02021, LPI Educational-Scientific Complex, grant of the President of the Russian Federation NSh-3796.2014.2.